\documentclass[11pt,a4paper]{article}

\usepackage[T1]{fontenc}
\usepackage{color}
\usepackage{graphicx}
\usepackage{mathtools}
\usepackage{ragged2e}

\usepackage{dcolumn}
\usepackage{bm}
\usepackage{graphicx,empheq}
\usepackage{verbatim} 
\usepackage[english]{babel}
\usepackage{hyperref}

\hypersetup{
citecolor=red,
colorlinks=true,
filecolor=red,
linkcolor=blue,
linktocpage=true,
urlcolor=blue
} 
\usepackage[titletoc,toc,title]{appendix}
\numberwithin{equation}{section}
\usepackage{cleveref}
\usepackage{epsfig}
\usepackage{amsfonts}
\usepackage{enumitem}

\newcommand\trick[1]{}
\newcommand{\be}{\begin{equation}} 
\newcommand{\ee}{\end{equation}}
\newcommand{\eq}[1]{(\ref{#1})}
\newcommand{\bit}{\begin{itemize}}  \newcommand{\eit}{\end{itemize}}
\newcommand{\ben}{\begin{enumerate}}  \newcommand{\een}{\end{enumerate}}

\newcommand{\bra}[1]{\langle #1|}
\newcommand{\ket}[1]{|#1 \rangle}

\newcommand{\rf}[1]{(\ref{#1})}

\def\bd{\begin{document}}
\def\ed{\end{document}}
\def\bea{\begin{eqnarray}}
\def\eea{\end{eqnarray}}
\let\bm=\bibitem

\def\la{\langle}
\def\ra{\rangle}

\def\npb#1#2#3{Nucl. Phys. {\bf{B#1}} #3 (#2)}
\def\plb#1#2#3{Phys. Lett. {\bf{#1B}} #3 (#2)}
\def\prl#1#2#3{Phys. Rev. Lett. {\bf{#1}} #3 (#2)}
\def\prd#1#2#3{Phys. Rev. {D bf{#1}} #3 (#2)}
\def\cmp#1#2#3{Comm. Math. Phys. {\bf{#1}} #3 (#2)}
\def\cqg#1#2#3{Class. Quantum Grav. {\bf{#1}} #3 (#2)}
\def\nppsa#1#2#3{Nucl. Phys. B (Proc. Suppl.) {\bf{#1A}}#3 (#2)}
\def\ap#1#2#3{Ann. of Phys. {\bf{#1}} #3 (#2)}
\def\ijmp#1#2#3{Int. J. Mod. Phys. {\bf{A#1}} #3 (#2)}
\def\rmp#1#2#3{Rev. Mod. Phys. {\bf{#1}} #3 (#2)}
\def\mpla#1#2#3{Mod. Phys. Lett. {\bf A#1} #3 (#2)}
\def\jhep#1#2#3{J. High Energy Phys. {\bf #1} #3 (#2)}
\def\atmp#1#2#3{Adv. Theor. Math. Phys. {\bf #1} #3 (#2)}

\def\sst{\scriptscriptstyle}
\def\thetabar{\bar\theta}
\def\Tr{{\rm Tr}}
\def\one{\mbox{1 \kern-.59em {\rm l}}}

\def\a{\alpha}      \def\da{{\dot\alpha}}  \def\dA{{\dot A}}
\def\b{\beta}       \def\db{{\dot\beta}}
\def\g{\gamma}  \def\G{\Gamma}  \def\dc{{\dot\gamma}}
\def\d{\delta}  \def\D{\Delta}  \def\ddt{\dot\delta}
\def\e{\epsilon}
\def\ve{\varepsilon}
\def\uve{\upvarepsilon}
\def\f{\phi}    \def\F{\Phi}    \def\vvf{\f}
\def\vphi{\varphi}
\def\h{\eta}
\def\k{\kappa}
\def\l{\lambda} \def\L{\Lambda}
\def\m{\mu} \def\n{\nu}
\def\o{\omega}
\def\p{\pi} \def\P{\Pi}
\def\r{\rho}
\def\s{\sigma}  \def\S{\Sigma}
\def\t{\tau}
\def\th{\theta} \def\Th{\Theta} \def\vth{\vartheta}
\def\X{\Xeta}
\def\z{\zeta}

\def\na{\nabla}

\def\cA{{\cal A}} \def\cB{{\cal B}} \def\cC{{\cal C}}
\def\cD{{\cal D}} \def\cE{{\cal E}} \def\cF{{\cal F}}
\def\cG{{\cal G}} \def\cH{{\cal H}} \def\cI{{\cal I}}
\def\cJ{{\mathscr J}} \def\cK{{\cal K}} \def\cL{{\cal L}}
\def\cM{{\cal M}} \def\cN{{\cal N}} \def\cO{{\cal O}}
\def\cP{{\cal P}} \def\cQ{{\cal Q}} \def\cR{{\cal R}}
\def\cS{{\cal S}} \def\cT{{\cal T}} \def\cU{{\cal U}}
\def\cV{{\cal V}} \def\cW{{\cal W}} \def\cX{{\cal X}}
\def\cY{{\cal Y}} \def\cZ{{\cal Z}}
\def\ct{{\cal t}}

\def\ua{\underline{\alpha}}
\def\uc{\underline{\phantom{\alpha}}\!\!\!\gamma}
\def\um{\underline{\mu}}
\def\ud{\underline\delta}
\def\ue{\underline\epsilon}
\def\una{\underline a}\def\unA{\underline A}
\def\unb{\underline b}\def\unB{\underline B}
\def\unc{\underline c}\def\unC{\underline C}
\def\und{\underline d}\def\unD{\underline D}
\def\une{\underline e}\def\unE{\underline E}
\def\unf{\underline{\phantom{e}}\!\!\!\! f}\def\unF{\underline F}
\def\unm{\underline m}\def\unM{{\underline M}}
\def\unn{\underline n}\def\unN{{\underline N}}
\def\unp{\underline{\phantom{a}}\!\!\! p}\def\unP{\underline P}
\def\unq{\underline{\phantom{a}}\!\!\! q}
\def\unQ{\underline{\phantom{A}}\!\!\!\! Q}
\def\unH{\underline{H}}

\def\As {{A \hspace{-6.4pt} \slash}\;}
\def\bs {{b \hspace{-6.4pt} \slash}\;}
\def\Ds {{D \hspace{-6.4pt} \slash}\;}
\def\Gts {{\Gt \hspace{-6.4pt} \slash}\;}
\def\ds {{\del \hspace{-6.4pt} \slash}\;}
\def\ss {{\s \hspace{-6.4pt} \slash}\;}
\def\ks {{ k \hspace{-6.4pt} \slash}\;}
\def\ps {{p \hspace{-6.4pt} \slash}\;}
\def\xs {{x \hspace{-6.4pt} \slash}\;}
\def\pas {{{p_1} \hspace{-6.4pt} \slash}\;}
\def\pbs {{{p_2} \hspace{-6.4pt} \slash}\;}
\def\cFs {{{\cal F} \hspace{-6.4pt} \slash}\;}
\def\Dss {{D \hspace{-7.5pt} \slash}\;}
\def\dss {{\del \hspace{-7.0pt} \slash}\;}

\def\Ah{{\hat{A}}}
\def\Dh{{\hat{D}}}
\def\Gh{{\hat{G}}}
\def\Fh{{\hat{F}}}
\def\Ih{{\hat{I}}}
\def\Jh{{\hat{J}}}
\def\Kh{{\hat{K}}}
\def\Lh{{\hat{L}}}
\def\Ph{{\hat{P}}}
\def\Rh{{\hat{R}}}
\def\Vh{{\hat{V}}}
\def\Xh{{\hat{X}}}
\def\Yh{{\hat{Y}}}

\def\ah{{\hat{\a}}}
\def\bh{{\hat{\b}}}
\def\gh{{\hat{\g}}}
\def\dh{{\hat{\d}}}
\def\rh{{\hat{\r}}}
\def\hh{\hat{h}}
\def\uh{\hat{u}}
\def\xh{\hat{x}}
\def\yh{\hat{y}}
\def\ph{\hat{p}}
\def\xih{\hat{\xi}}
\def\chih{\hat{\chi}}
\def\Psih{\hat{\Psi}}
\def\phih{\hat{\phi}}

\def\psit{\tilde{\psi}}
\def\Psit{\tilde{\Psi}}
\def\Psibt{\tilde{\bar{Psi}}}

\def\lambdat{\tilde {\lambda}}
\def\st{\tilde{\sigma}}

\def\delt{\tilde{\delta}}
\def\Phit{\tilde{\Phi}}
\def\Phitb{\overline{\tilde{Phi}}}
\def\tht{\tilde{\th}}
\def\lt{\tilde{\l}}
\def\chit{\tilde{\chi}}
\def\phit{\tilde{\phi}}

\def\At{\tilde{A}}
\def\Bt{\tilde{B}}
\def\Ct{\tilde{C}}
\def\Dt{\tilde{D}}
\def\Et{\tilde{E}}
\def\Ft{\tilde{F}}
\def\Gt{\tilde{G}}
\def\Ht{\tilde{H}}
\def\It{\tilde{I}}
\def\Jt{\tilde{J}}
\def\Qt{\tilde{Q}}
\def\Rt{\tilde{R}}
\def\Mt{\tilde{M }}
\def\Nt{\tilde{N}}
\def\St{\tilde{S}}
\def\Vt{\tilde{V}}
\def\Xt{\tilde{X}}
\def\at{\tilde{a}}
\def\dt{\tilde{d}}
\def\htt{\tilde{h}}
\def\ft{\tilde{f}}
\def\gt{\tilde{g}}
\def\pt{\tilde{p}}
\def\qt{\tilde{q}}
\def\vt{\tilde{v}}
\def\nt{\tilde{n}}
\def\ut{\tilde{u}}
\def\wt{\tilde{w}}
\def\zt{\tilde{z}}
\def\xt{\tilde{x}}
\def\yt{\tilde{y}}
\def\Psit{\tilde{\Psi}}
\def\vphit{\tilde{\varphi}}
\def\tD{\tilde{\D}}

\def\eb{\bar{\epsilon}}
\def\delb{\bar{\partial}}
\def\thb{\bar{\theta}}
\def\mub{\bar{\mu}}
\def\lamb{\bar{\l}}
\def\psib{\bar{\psi}}
\def\sb{\bar{\sigma}}
\def\xib{\bar{\xi}}
\def\chib{\bar{\chi}}

\def\Psib{\bar{\Psi}}
\def\Phib{\bar{\Phi}}
\def\Lamb{\bar{\Lambda}}
\def\Sb{{\overline \Sigma}}
\def\cb{\bar{c}}
\def\hb{\bar{h}}
\def\qb{\bar{q}}
\def\wb{\bar{w}}
\def\ub{\bar{u}}
\def\zb{{\bar{z}}}
\def\Hb{\bar{H}}
\def\Qb{{\bar Q}}
\def\Omegab{\overline{\Omega}}
\def\ob{\overline{\omega}}

\def\Ab{{\overline A}} \def\Bb{{\overline B}} \def\Cb{{\overline C}}
\def\Db{{\overline D}} \def\Eb{{\overline E}} \def\Fb{{\overline F}}
\def\Gb{{\overline G}}
\def\Ib{{\overline I}}
\def\Jb{{\overline J}} \def\Kb{{\overline K}} \def\Lb{{\overline L}}
\def\Mb{{\overline M}} \def\Nb{{\overline N}} \def\Ob{{\overline O}}
\def\Pb{{\overline P}}  \def\Rb{{\overline R}}
 \def\Tb{{\overline T}} \def\Ub{{\overline U}}
\def\Vb{{\overline V}} \def\Wb{{\overline W}} \def\Xb{{\overline X}}
\def\Yb{{\overline Y}} \def\Zb{{\overline Z}}

\def\fb{{\overline f}}
\def\gb{{\overline g}}
\def\nb{{\overline n}}
\def\mb{{\overline m}}
\def\lb{{\overline l}}
\def\yb{{\overline y}}

\def\ldel{{\overleftarrow{\del}}}
\def\rdel{{\overrightarrow{\del}}}
\def\ldeldel{{\overleftarrow{\del^2}}}
\def\rdeldel{{\overrightarrow{\del^2}}}
\def\ldelb{{\overleftarrow{\bar{\del}}}}
\def\rdelb{{\overrightarrow{\bar{\del}}}}

\def\ba{{\bf a}}
\def\bk{{\bf k}}
\def\bl{{\bf l}}
\def\bp{{\bf p}}
\def\bq{{\bf q}}
\def\br{{\bf r}}
\def\bt{{\bf t}}
\def\bu{{\bf u}}
\def\bv{{\bf v}}
\def\bx{{\bf x}}
\def\by{{\bf y}}
\def\bA{{\bf A}}
\def\bR{{\bf R}}
\def\bV{{\bf V}}

\def\bz{{\boldsymbol{\zeta}}}

\def\bone{{\bf 1}}

\def\va{{\vec a}}
\def\vk{{\vec k}}
\def\vp{{\vec p}}
\def\vq{{\vec q}}
\def\vx{{\vec x}}
\def\vy{{\vec y}}
\def\vu{{\vec u}}
\def\vv{{\vec v}}
\def \vH{{\vec H}}
\def \vg{{\vec g}}

\def\vs{{\vec \sigma}}
\def\vtau{{\vec \tau}}

\newcommand{\ov}[1]{\overrightarrow{#1}}

\def\frA{\mathfrak{A}}
\def\frB{\mathfrak{B}}
\def\frC{\mathfrak{C}}
\def\frD{\mathfrak{D}}
\def\frE{\mathfrak{E}}
\def\frF{\mathfrak{F}}
\def\frG{\mathfrak{G}}
\def\frH{\mathfrak{H}}
\def\frM{\mathfrak{M}}
\def\frN{\mathfrak{N}}
\def\frR{\mathfrak{R}}
\def\frW{\mathfrak{W}}

\def\fra{\mathfrak{a}}
\def\frb{\mathfrak{b}}
\def\frf{\mathfrak{f}}
\def\frg{\mathfrak{g}}
\def\frh{\mathfrak{h}}
\def\frl{\mathfrak{l}}
\def\frs{\mathfrak{s}}
\def\fri{\mathfrak{i}}
\def\frj{\mathfrak{j}}

\def\ma{\mathfrak{a}}
\def\mg{\mathfrak{g}}
\def\mh{\mathfrak{h}}
\def\mR{\mathfrak{R}}
\def\mN{\mathfrak{N}}

\newcommand{\nn}{{\nonumber}}

\def\d{\delta}\def\D{\Delta}\def\ddt{\dot\delta}

\def\pa{\partial} \def\del{\partial}
\def\xx{\times}
\def\uno{\mbox{1 \kern-.59em {\rm l}}}

\def\trp{^{\top}}
\def\inv{^{-1}}
\def\dag{\dagger}
\def\pr{^{\prime}}

\def\rar{\rightarrow}
\def\lar{\leftarrow}
\def\lrar{\leftrightarrow}

\newcommand{\0}{\,\!}      
\def\one{1\!\!1\,\,}
\def\im{\imath}
\def\jm{\jmath}

\newcommand{\tr}{\mbox{tr}}
\newcommand{\slsh}[1]{/ \!\!\!\! #1}

\newcommand{\1}{\mbox{1}\hspace{-0.25em}\mbox{l}}

\def\vac{|0\rangle}
\def\lvac{\langle 0|}

\def\hlf{\frac{1}{2}}
\def\ove#1{\frac{1}{#1}}
\newcommand{\hot}[1]{\frac{#1}{2}}

\def\Box{\square}
\def\CC {\mathbb{C}}
\def\FF {\mathbb{F}}
\def\RR{\mathbb{R}}
\def\NN{\mathbb{N}}
\def\ZZ{\mathbb{Z}}
\def\bb#1{{\bf #1}}
\def\bcomment#1{}
\def\bfhat#1{{\bf \hat{#1}}}
\def\VEV#1{\left\langle #1\right\rangle}

\newcommand{\ex}[1]{{\rm e}^{#1}} \def\ii{{\rm i}}

\newcommand{\lrbrk}[1]{\left(#1\right)}
\newcommand{\lrsbrk}[1]{\left[#1\right]}
\newcommand{\sfrac}[2]{{\textstyle\frac{#1}{#2}}}

\def\stw{{\sqrt{2}}}

\def\rf {{\rm f}}
\def\ri {{\rm i}}
\def\rj {{\rm j}}
\def\rn {{\rm n}}
\def\rk {{\rm k}}
\def\rl {{\rm l}}
\def\rr {{\rm r}}
\def\rs {{\scriptscriptstyle \rm S}}
\def\rt {{\scriptscriptstyle \rm T}}

\def\rQ {{\scriptscriptstyle \rm \cQ}}
\def\rR {{\scriptscriptstyle \rm \cR}}

\def\cQb{{\cal \Qb}}
\def\cRb{{\cal \Rb}}
\def\cWb{{\cal \Wb}}

\def\fd {{\rm N}}
\def\afd {{\overline{\rm N}}}

\def \II {I\hspace{-.1em}I\hspace{.1em}}
\def \IIA {\mbox{\II A\hspace{.2em}}}
\def \IIB {\mbox{\II B\hspace{.2em}}}
\def \gs {g^s}
\def \ls {\lambda^s}

\def \I {{\cal I}}
\def \qs {q\hspace{-.53em}/\hspace{.15em}}
\def \ks {k\hspace{-.53em}/\hspace{.15em}}
\def \YM {{\mbox{\tiny YM}}}
\def \gym {g_{\YM}}

\def \Lc {\L_c}
\def\IR{\relax{\rm I\kern-.18em R}}
\def \id {{\bf 1}}

\def\cci{\ell}
\def\ccj{\ell'}

\def\bbq{\pmb{q}}
\def\bom{\pmb{\o}}
\def\bJ{\pmb{J}}
\def\bM{\pmb{M}}
\def\bB{\pmb{B}}
\def\bn{\pmb{n}}
\def\bE{\pmb{E}}

\newcommand{\rrr}[1]{\vskip 0.2cm \noindent{\bf #1} ---}

\long\def\symbolfootnote[#1]#2{\begingroup%
\def\thefootnote{\fnsymbol{footnote}}\footnote[#1]{#2}\endgroup}
\long\def\RemarkBox#1{\begin{flushleft}\fbox{\begin{minipage}
{17.5cm}{\bf Remark:} ~#1\end{minipage}}\end{flushleft}}
\newcommand{\aei}{\it Max Planck Institute for Gravitational Physics
(Albert Einstein Institute)\\ Am M\"uhlenberg 1, 14476 Golm,
Germany}

\newcommand{\nthu}{{\it Department of Physics, National Tsing-Hua
  University,
  Hsinchu 30013, Taiwan}}

\newcommand{\ctc}{{\it
Center of Theory and Computation, 
National Tsing-Hua University, Hsinchu 30013, Taiwan}}

\newcommand{\ncts}{{\it
National Center for Theoretical Sciences, Taipei 10617, Taiwan}}

\begin{document}
\begin{center}
\thispagestyle{empty}
              {\Large \bf A Matrix Model Proposal for Quantum Gravity and the \\
                Quantum Mechanics of  Black Holes}
           
\vspace{25pt}
Chong-Sun Chu

\vspace{5pt}\nthu
\\
\vspace{5pt}\ctc
\\
\vspace{5pt}\ncts
\vspace{0.5cm}


\begin{abstract} We propose a quantum mechanical
  theory of quantum spaces
  described by large $N$ noncommutative geometry
  as a model for quantum gravity.  The
  model
  admits fuzzy sphere as static solution.  Over the fuzzy
  geometry, the quantum mechanics of the fermions is given by a sum of
  oscillators with equal frequency. The energy state where exactly
  half of the Fermi sea is filled contains the maximal amount of
  degeneracy. This state of the fuzzy sphere obeys the mass-radius
  relation of a Schwarzschild black hole if the fuzzy sphere is
  identified with the black hole horizon. Moreover the set of states
  in the Fermi sea gives precisely the Bekenstein-Hawking entropy. We
  thus propose that quantum black holes are described by fuzzy spheres
  with a half-filled Fermi sea in our
  model.
 We also consider a system of two fuzzy spheres by embedding them as
 blocks in the matrix quantum mechanics.  When the distance $r$
 between the two fuzzy spheres is small, the total energy of the system
 can be computed using perturbation theory.  We show that in the
 leading order of large $N$ limit, the interaction energy depends on
 $- G M_1 M_2$ exactly
 the manner as
 in Newton gravity. To
 reproduce the correct $r$ dependence in the long
 range,  we expect the inclusion of
 large $N$ corrections and quantum effects will be  needed.

\end{abstract}

\end{center}

\newpage

\setcounter{page}{1}


\section{Introduction}

Black hole arises as
solution of the classical general relativity. However, a number of
properties of black hole, such as the existence of an area dependent
Bekenstein-Hawking entropy \cite{Bekenstein:1972tm,Bekenstein:1973ur},
the black hole information puzzle
\cite{Hawking:1976ra,Polchinski:2016hrw,Almheiri:2020cfm},
the existence of a black hole singularity
\cite{penrose:1964wq},
the exhibition of holography \cite{tHooft:1993dmi,Susskind:1994vu} etc,  
are puzzling and it is believed that only in a
theory of quantum gravity can these problems
be properly understood and resolved.

A prime candidate to a theory of quantum gravity is string theory. Previously,
brane models of black hole have been considered in
non-perturbative string theory, with the Berkenstein-Hawking entropy
reproduced successfully from a microstates counting
\cite{Strominger:1996sh,Mathur:2005zp}.  Recently progress has been
achieved for black holes constructed in the AdS/CFT correspondence
\cite{Maldacena:1997re} where the Page curve behavior of the
entanglement entropy of the Hawking radiation is remarkably obtained
\cite{Engelhardt:2014gca,Penington:2019npb,Almheiri:2019psf,Almheiri:2019hni}.
Nevertheless, despite these success, it is highly desirable to have a
direct formulation of the quantum gravity itself and be able to
describe the fundamental degrees of freedom of quantized spacetime and
the set of microstates explicitly without resorting to supersymmetry and
duality.  In this regards, large $N$ matrix model formulation of
string theory, such as the BFSS matrix model \cite{Banks:1996vh}
or the IKKT matrix model \cite{Ishibashi:1996xs},
is particularly appealing since in matrix model, space/time
emerges from the more fundamental quantum mechanical matrix degrees
of freedom in the large $N$ limit, and one can study the quantum
properties of space time using the matrix quantum mechanics.

The main stream approach to quantum black hole has been a top-down one
where the properties of the black hole is studied using a certain candidate
theory of quantum gravity such as those mentioned above.
However so far it has not been possible to study
a Schwarzschild black hole this way such that the expected properties
of a quantum black hole can be obtained and addressed.
In a recent series of studies \cite{Chu:2022ieq,Chu:2023mqi,Chu:2023agv}, we
have initiated a bottom-up approach to quantum black hole by 
using the anticipated properties of quantum black hole as ``empirical
input'' to guide the construction of the fundamental theory of quantum
gravity.  We took the assumption that quantum gravity can be
formulated in terms of a quantum mechanics of fermionic and bosonic
degrees of freedom, and a generic  quantum mechanical model of black hole was
proposed. The degrees of freedom of the quantum mechanics is supposed
to take some generic form
\be \label{matrix model}
L= i \psi^+ \dot{\psi}+  \psi^+ h(X) \psi-V(X),
\ee 
where $ h(X) $ denotes some Yukawa coupling and $V(X)$ the
self-interaction.
Our goal was to identify basic
properties of quantum gravity which are essential to the
construction of the theory. Our hope was that the model would be helpful
in ways similar to that of the Bohr atomic model to the building of
quantum mechanics.
We found \cite{Chu:2023mqi} that if our model admits a constant density
of energy eigenstates and if the Fermi sea of the system is filled up
to a Fermi energy level that is inversely proportional to the system
size, then the Schwarzschild radius of black hole is reproduced for
the system.  Moreover the system is in a highly degenerate energy state whose
counting of microstates gives precisely  the Bekenstein-Hawking
entropy. While  the result is quite encouraging,
the identified conditions are sufficient ones, and 
it is not clear if all the assumptions can indeed be implemented 
consistently in a quantum mechanics.
The construction of a consistent
quantum
mechanical model  of spacetime which includes quantum black hole
is the main motivation of this work.

In this work, we propose an explicit large $N$ quantum mechanics of
non-abelian bosonic and fermionic variables as a model of quantum
gravity.
The use of large $N$ is
inspired by the previous success of the BFSS and IKKT matrix models,
see, for example, \cite{Taylor:2001vb,Ydri:2017ncg} for review.
There are three bosonic variables $X^a$ in the adjoint of
$SU(N)$ which correspond to the three spatial coordinates of a
quantized
noncommutative space. We
note that the emergence of noncommutative geometry is generally
expected 
when space (or spacetime) is quantized in quantum gravity
\footnote{The studied of quantized spacetime in terms of noncommutative
geometry goes back to the study of Snyder \cite{Snyder:1946qz}
and Yang \cite{Yang:1947ud}.
}.
In addition, there are two fermionic variables
which couple minimally to the geometry, and make up 
the fundamental Hilbert space of quantum gravity.
We show that our
model
of quantized spacetime
admits fuzzy sphere as static
  solution.  Over the fuzzy geometry, the quantum mechanics of the
  fermions gives a Fermi sea with uniform energy levels
  in the large $N$ limit.
  It is amazing that the energy state where exactly half of the Fermi sea is
  filled contains the maximal amount of degeneracy and satisfies the
  the mass-radius relation of a Schwarzschild black
  hole if the fuzzy sphere is identified with the black hole
  horizon. Moreover the set of states in the half-filled
  Fermi sea gives precisely the Bekenstein-Hawking entropy.
  We therefore propose that quantum black holes
  are described by fuzzy spheres with a half-filled Fermi sea in our
  model.
  Quite  amazingly, the quantum
mechanics proposed here  realizes pretty much all the ideas outlined
earlier in \cite{Chu:2022ieq,Chu:2023mqi,Chu:2023agv}. 

To show that the
fuzzy spheres are gravitating objects, one needs to show that the
  interaction between them agrees with gravity, e.g. with
  Newton gravity in the long distance limit. We thus
  consider a system of two fuzzy spheres by embedding them as
 blocks in the matrix quantum mechanics.  When the distance $r$
 between the two fuzzy sphere is small, the total energy of the system
 can be computed using perturbation theory.  We show that in the
 leading order of large $N$ limit, the interaction energy depends on
 the product of black holes masses and the Newton constant exactly as is
 expected in Newton gravity.
 it is possible that the inclusion of
 large $N$ corrections as well as quantum loop effects
 could reproduce the  correct Newton limit
 at large distance. We leave this important problem for further study.

 The plan of the paper is as follows. In section 2, we give our proposal of
 a large $N$ quantum mechanics as a fundamental formulation of
 quantum gravity in 3-dimensions. In section 3, we show that the fuzzy
 sphere solution with a half-filled Fermi sea has the desired
 properties of a quantum black hole, namely with the black hole
 mass-radius relation and the Bekenstein-Hawking black hole entropy
 reproduced correctly.In section 4, we discuss the mode stability of our fuzzy
 sphere solution and show that it is stable with respect to mass preserving
 perturbations.
 In section 5, we consider a system of 2
 fuzzy spheres and derive their interaction energy in the small separation
 limit. We discuss how the inclusion of
 large $N$ corrections as well as quantum loop effects
 could generate the desired gravitational forces between the quantum black
 holes. Further discussion is found
 in section 6.

\section{A proposed model of quantum gravity}

Let us consider a
model of quantized space where the spatial coordinates  
becomes operators $X^a$ ($a=1,2,3$)
and are represented by $SU(N)$ matrices. Here $N$ is 
large and characterizes the dimensions of the Hilbert space of quantum gravity.
In addition, we propose to include fermionic degrees of freedom in
the fundamental formulation of quantum gravity. In our  model proposed below,
they are given by the 2-components
spinors $\psi^\dagger, \psi$ in the adjoint representation
of $SU(N)$.
Our choice of degrees of freedom is a minimalistic one and is allowed since
we do not assume supersymmetry. In fact, 
it is important to realize that going bottom-up,
there is no physical reason to insist on having supersymmetry. We 
remark that in the usual top down approach, supersymmetry is often needed for
the reason of consistency of formulation.
e.g. for  string/M theory in 10 or 11 dimensions.
However, supersymmetry is irrelevant for the well-definedness  
and consistency of quantum mechanics as there is no issue of renormalizability
here.
Also the usual advantages of supersymmetric
quantum mechanics such as exact computation of quantities such as index or
partition function does not appeal to us since here we are after
the construction
of a physical theory rather than a theory that
is exact or calculable. Therefore we will not insist on having supersymmetry.
This allows for a  much wider choice of terms in the Lagrangian which
are forbidden otherwise.

We propose to consider the following
quantum
mechanical model of spacetime, with the Lagrangian
\be \label{L}
L = \tr \left[
\frac{1}{2 M_0 } \dot{X}^{a2} 
+\frac{M_P}{N^2} \left(
[X^a,X^b]^2
+ 4 X^{a2} \right)
+ i \dot{\psi}^\dag \psi
- a_2  \frac{M_P}{N^2}  \psi^\dag \s^a X^a  \psi   \right] - a_3 r_X M_P 
\ee
and the Hamiltonian
\be \label{H}
H = \tr\left[
  \frac{M_0}{2} P^{a 2}
  -\frac{M_P}{N^2} \left( [X^a,X^b]^2 + 4 X^{a2} \right)
+ a_2\frac{M_P}{N^2} \psi^\dag \s^a X^a  \psi \right] + a_3 r_X M_P .
\ee
where $P^a = \dot{X}^a/M_0$ is  the conjugate momentum for the bosonic
degrees of freedom.  The  matrices
  $X^a,  \psi, \psi^\dag$ are  dimensionless $N \times N$ traceless
  Hermitian matrices, meaning that the model \eq{L}
  is supposed to describe the dynamics
  with respect to the center of mass frame.
  $\s^a$ are the Pauli matrices.
 We note  that unlike the IKKT or BFSS matrix
model where the maximal supersymmetry fixes the field content
and the action uniquely, our model is nonsupersymmetric and there is a negative
mass term.  In \eq{L},
the quartic term is fixed to have a coefficient 4 times that of the
mass term. This can always be achieved with a
rescaling of the variables $X^a$. Here  $M_0 = a_0^2 M_P$ and
the Lagrangian is specified by a   mass scale $M_P$ (Planck mass) that sits
in front of the bosonic potential, whose
relation with the Newton constant $G$ will be specified below.
The field $\psi$ is normalized so
that the fermionic kinetic term has a unit coefficient. We have adopted a
definite $N^2$ dependence in the bosonic potential and the Yukawa coupling term.
The bosonic kinetic term and the Yukawa term are then specified up to a choice
of the coefficients $a_0$ and $a_2$.
In addition, we have included a topological action term specified by $a_3$.
Here $r_X$ is the rank of the matrix $\G:= [X^a, X^b]^2$. The $r_X$ term
does not affect the equation of motion, but measures the energy of space
due to noncommutativity.  It is $r_X =0$ for  Abelian configurations
(including the Minkowski vacuum $X^a =0$)
and $r_X = N$ for the fuzzy sphere  solution.
We note that the
coefficients $a_0, a_2, a_3$ are dimensionless. They are not fixed by
symmetry argument, but will be determined below
``phenomenologically'' by requiring the properties of the
fuzzy sphere solutions match up
with that expected of the quantum Schwarzschild black holes. 
In  the large $N$ limit,  our model is equivalent to some
non-supersymmetric version of the IKKT-like instantonic model in 4 dimensions
by a large $N$ reduction \cite{Eguchi:1982nm}. 

The
Lagrangian \eq{L} has an $SO(3)$ rotational invariance where $X^i$ transforms as
a vector and $\psi$ in the spinor representation.
The theory is also  invariant under the global $SU(N)$ transformation
\be
X^a \to U X^a U^\dag, \quad \psi \to U \psi U^\dag, \quad
\psi^\dag \to U \psi^\dag U^{\dag},
\qquad U U^\dag =1. 
\ee
We remark that in the SUSY BFSS matrix model, the $SU(N)$ symmetry
is usually gauged in order to close the SUSY algebra. In
\cite{Maldacena:2018vsr}, it was shown that the global BFSS matrix QM,
although it contains extra non-singlet states that do
not come in supersymmetry multiplets, is nevertheless also consistent.
In our case, it is also possible
to  gauge the $SU(N)$ and impose
a singlet Gauss law constraint. We do not consider this possibility here
but simply noting that the fuzzy sphere solution is also a solution of
the gauged
model and our analysis below remains the same. 
%
We remark that our proposal
is similar in philosophy to that of the 
M(atrix) theory \cite{Taylor:2001vb}. 
However we do not assume supersymmetry, and so we can
write down our model directly for 3 space dimensions, and we can include
a mass term. 
We note  that a mass term has been considered in the IKKT matrix model 
where interesting classical solutions such as fuzzy spacetime
\cite{Steinacker:2021yxt}  and
expanding universe \cite{Kim:2011ts,Kim:2012mw, steinacker:2017vqw,
Sperling:2019xar}, 
have been obtained.
However, the possible connection of
a nonsupersymmetric large $N$ quantum mechanics
with black hole is new.

\section{Fuzzy sphere solution with Fermi sea}

The classical equation of motion for a bosonic matrix configuration 
is given by
   \be \label{eom1}
   - \frac{1}{M_0}\ddot{X}^a +\frac{4  M_P}{N^2}
   \left( [X^b,[X^a,X^b]] + 2 X^a \right) =0, \quad \psi =0.
\ee
For static configuration, this becomes
\be
[[X^a,X^b],X^b] = 2 X^a.
\ee
This can be solved by the spin $j =(N-1)/2$ representation of
$SU(2)$, which are given by $2j+1=N$ dimensional matrices satisfying
\be\label{FS1}
         [X^a,X^b] = i \e_{abc} X^c.
\ee
Due to the Casimir relation 
\be \label{FS2}
\sum_a X^{a2} = \frac{N^2-1}{4} \id
\ee
the configuration \eq{FS1}, \eq{FS2} define a fuzzy sphere.
Since $r_X=N$, the classical fuzzy sphere solution
carries an energy
\be\label{E_B}
E_B = (2a_3 - 1)\frac{M_P N}{2}
\ee
in the leading order limit of large $N$.

Next, let us analysis the Yukawa coupling term
\be
H_F = a_2  \frac{M_P}{N^2}  \psi^\dag \s^a X^a  \psi .
\ee
Over the  fuzzy sphere geometry, the matrix $K := \s^a X^a$ satisfies
$K^2 + K - \frac{N^2-1}{4} \id =0$ and so it has eigenvalues $(N-1)/2$ 
or $-(N+1)/2$. Since it 
is traceless, therefore  $N+1$ of the eigenvalues are positive and
$N-1$ of them are the negative ones.  Let us from now on
  consider the leading large $N$ limit. In the leading order of large $N$ limit,
$K$ has the eigendecomposition 
\be \label{K-eigen}
K_{(m\a) (n\b)} = \frac{N}{2} \; \sum_{p=1}^{N}
\left(\cU^{p}_{m\a} \cU^{p\dag}_{n\b} - \cV^{p}_{m\a} \cV^{ p \dag}_{n\b}
  \right).
  \ee
Here $\cU^{p}_{n\b}, \cV^{p}_{n\b}, p =1, \cdots, N$
  are eigenvectors of $K$ with positive
  and negative  eigenvalue.
  Introduce  the fermionic oscillators
  \be \label{xi-chi}
  \xi^{p}_k := \cU^{p \dag}_{n\b} \psi_{nk\b}, \quad
  \chi^{p\dag}_k := \cV^{p \dag}_{n\b} \psi_{nk\b}
  \ee
  and
$H_F$ reads
   \be \label{HF-diag-classical}
   H_F = \frac{a_2 M_P}{2N}
  \sum_{p,k=1}^N    \left(
     \xi_k^{p\dag} \xi_k^{p} - \chi_k^{p} \chi_k^{p\dag} \right).
     \ee
     Here the   Grassmanian variables $\xi_k^{p}$ and  $\chi_k^{p}$
     resemble the oscillators $b$ and $d$ of the Dirac theory of electrons.

The theory    \eq{H} can be  canonically quantized. For the
bosonic variables, we impose the commutation relation
\be
   [X^a_{mk}, P^b_{nl} ] = i \d^{ab} \d_{mn} \d_{kl}.
   \ee
As for  the fermionic variables, the conjugate momentum $\pi = i \psi^\dag$
give rises to the anti-commutation relation
\be \label{psid-psi}
   \{\psi^\dag_{mk \a}, \psi_{nl \b}\} = \d_{mn} \d_{kl} \d_{\a\b}.
   \ee
This gives equivalently,    
\be
\{\xi^{p}_k , \xi^{q \dag}_l \} = \d^{pq} \d_{kl}, \quad
\{\chi^{p}_k , \chi^{q\dag}_l \} = \d^{pq} \d_{kl}.
\ee
The quantized fermion Hamiltonian can be obtained from \eq{HF-diag-classical}
with the subtraction of a constant,
\be\label{HF-diag}
H_F = \frac{a_2 M_P}{2N}
  \sum_{p,k=1}^N    \left(
     \xi_k^{p\dag} \xi_k^{p} + \chi_k^{p \dag} \chi_k^{p} \right)
     \ee
such that there is no zero point energy for the oscillators.
This may also be obtained with a prescription of
normal ordering defined on the oscillators
$\xi^{p}_k$ and $\chi^{p}_k$,
$:\chi^{p}_k \chi^{p\dag}_k: = - \chi^{p\dag}_k \chi^{p}_k$ etc.
The quantized  Hamiltonian
$H_F$ has the eigenstates
\be \label{es}
\left|\Psi^{p_1 \cdots p_r q_1 \cdots q_s}_{k_1 \cdots k_r l_1 \cdots l_s}\right\rangle
:=\xi^{p_1\dag}_{k_1}\cdots \xi^{p_r\dag}_{k_r} \chi^{q_1\dag}_{l_1} \cdots
\chi^{q_s\dag}_{l_s} \ket{0},
\ee
where $\ket{0}$ is the fermionic Fock vacuum defined by
\be
\xi^p_k \ket{0} = \chi^q_l \ket{0} =0, \quad \forall p,q,k,l
\ee
and the eigenvalues
\be \label{HF}
E_F = \frac{a_2 M_P}{2N} (N^2 + n), \quad n := r+s-N^2,
\ee
where $n = -N^2, \cdots,  N^2$
specifies the energy level within the Fermi sea.
The lowest level $n=-N^2$ corresponds
to an empty Fermi sea,
while the highest level $n =N^2$ corresponds to an
completely filled Fermi sea. 
We remark that one may label  the oscillators with the collective indices
$a := (p,k)$ and the fermionic Hamiltonian can be written as
\be \label{H-cells}
H_F =
\frac{a_2 M_P}{2N}
     \sum_{a=1}^{N^2} \xi_a^\dag \xi_a + \chi_a^\dag \chi_a.
     \ee
 Physically,  \eq{H-cells}  may be given an
interpretation that     the fuzzy sphere has been
     divided into $N^2$ cells with unit cell area of $\Delta A = 4 \pi l_P^2$
     of the Planck size.
Each cell, whose location is labeled by $a = (p,k)$,
is populated by a pair of oscillators $\xi_a, \chi_a$ which describes the
quantum fluctuations over the fuzzy sphere.
This is strikingly
similar to the description
of ``partons'' in the holographic picture of black hole
suggested by Susskind \cite{Susskind:1994vu}.
It is quite amusing that this conjectured
feature of holography get a concrete realization
in our quantum mechanics of spacetime.
In the next section, we propose to identify the quantum Schwarzschild
black hole with the fuzzy sphere solution with a half-filled Fermi sea $n=0$
in our model.

Before we move on, it is necessary to discuss the stability of the fuzzy sphere
solution. In general relativity, it is known that
the Schwarzschild metric and the Kerr metric
is stable against mode perturbations that do not change the mass $M$
and angular
momentum $J$ of these vacuum solutions. This statement of mode stability
was originally proven \cite{Regge:1957td,Whiting:1988vc}
for perturbations that can be written as a superposition of
spherical harmonic modes. Recently, it has been shown that
\cite{Dafermos:2016uzj}
Schwarzschild metric is stable against general linear perturbations
that does not necessarily need to be written
as a finite sum of spherical harmonic modes. However, the linear stability of
Kerr metric remains an open
problem. The full non-linear stability of vacuum solution in general relativity
is a hard problem and has only been established for the Minkowski metric
\cite{Christodoulou:1993uv}. For us here, we will be satisfied with a
mode analysis and
shows that our fuzzy sphere solution is stable against mode perturbations that
do not change the energy of the solution. 

For linearized analysis, we look at the
quadratic potential of the fluctuations $\d X^a$.
Over a general classical solution $X^a$, we have
\be \label{V-def}
U = \frac{ M_P}{N^2} V,
\quad
V = -\tr \left( F_{ab}^2 +2 [X^a,X^b][\d X^a, \d X^b] +4 \d X^a \d X^a \right),
\ee
where
\be
F_{ab} := [X^a, \d X^b] - [X^b, \d X^a]
\ee 
is the field strength.
The diagonalization of $V$ is standard.
It is convenient to introduce  the derivative operator $L^a$
whose action on a matrix $f$ is given by $L^a f = [X^a,f]$.
In terms of which we have
\be
F_{ab} = -i \e_{abc} (\ve \cdot L)_{cd} \d X^d,
\ee
where $\ve_a$ is a vector matrices with components $(\ve_a)_{bc} := -i \e_{abc}$.
So far this is general.
For  $X^a$ given by the fuzzy sphere \eq{FS1}, \eq{FS2},
the second term in $V$ of
\eq{V-def} can also be expressed in terms of $F_{ab}$ and
the potential takes the compact form
\be
V = 2 \tr (
  \d X^a
  N_{ab}
  \d X^b ), \quad \mbox{where} \quad
  N_{ab} :=  \left( 
(\ve \cdot L)^2 - (\ve \cdot L) -2 \right)_{ab}.
  \ee
As a result, the equation of motion for the linearized fluctuation is given by
\be \label{l-EOM}
\d \ddot{X}^a + \frac{4 a_0^2 M_P^2}{N^2} N_{ab} \d X^b =0,
\ee 
and the study of the mode stability of the fluctuations
reduces to the eigenvalue problem of  $N_{ab}$:
\be \label{Nphi}
N_{ab} \phi^b = \l \phi^a.
\ee

The operator $N_{ab}$ 
is diagonalized by the vector harmonics.  To construct them,  note that
$\ve_a$ is an angular momentum of angular momentum quantum number
$\ell_\ve =1$ since 
\be
[\ve_a, \ve_b] = i \e_{abc} \ve_c, \quad  \ve_a^2 = 2.
\ee
An eigenbasis is given by $\ket{\ell_\ve, m_\ve}$, $m_\ve = -1,0,+1$:
\be
\ve^2 \ket{\ell_\ve, m_\ve} = 2 \ket{\ell_\ve, m_\ve}, \quad
\ve_z \ket{\ell_\ve, m_\ve} = m_\ve \ket{\ell_\ve, m_\ve}.
\ee
Note also that for the fuzzy sphere background, $L^a$ is
an angular momentum operator.  An eigenbasis is given by
$\ket{\ell, m_z}$,
\be
L^2 \ket{\ell, m_z} = \ell (\ell+1)\ket{\ell, m_z}, \quad
L_z \ket{\ell, m_z} = m_z \ket{\ell, m_z}
\ee
for  $m_z =-\ell, \cdots, \ell$ and $0 \leq \ell \leq  N-1$.
Here the angular momentum quantum number $\ell$ is cut off by $N$
due to the noncommutativity of the fuzzy sphere. 
The spherical harmonics $\Yh^\ell_{m_z} $ is now obtained
in the matrix representation as
$(\Yh^\ell_{m_z})_{n_1 n_2} =
\langle n_1 n_2 | \ell, m_z \rangle$.
The eigenvalue problem of the operator $\ve \cdot L$ can now be
easily solved with the help of 
the total angular momentum operator
$J^a := \ve^a + L^a$.
$(J^2, J_z)$ is diagonalized by 
\be
J^2 \ket{J, M_J} = J(J+1) \ket{J, M_J}, \quad
J_z \ket{J, M_J} = m_J \ket{J, M_J},
\ee
and
\be \label{Jl}
\mbox{$J=1$ for $\ell =0$}
\quad \mbox{and}\quad
\mbox{ $J = \ell-1, \ell, \ell+1$ for $\ell \geq 1$}.
\ee
As a result,  $\ve \cdot L = \frac{1}{2} (J(J+1) - \ell (\ell+1) -2)$ and
$\l$ has the eigenvalues,
\be \label{eee}
\ve \cdot L = \begin{cases}
  0 &  \\
  -(\ell +1)  & \\
  -1 &  \\
  \ell &  \\
  \end{cases},
\quad
\l = \begin{cases}
  -2 & \qquad \mbox{for $\ell =0, J =1$}\\
   \ell (\ell +3)  & \qquad \mbox{for $\ell \geq 1, J =\ell -1$}\\
  0 &  \qquad \mbox{for $\ell \geq 1, J =\ell $}\\
  (\ell +1)(\ell-2) &  \qquad \mbox{for $\ell \geq 1, J =\ell +1$} \\
  \end{cases} .
\ee
For given $\ell$ and $J$ as given by \eq{Jl},
the corresponding eigenstate can be constructed by
using the CG coefficient  $C^{JM_J}_{\ell m_z 1 p}$ 
\be
\ket{J, M_J} = \sum_{m_z=-\ell}^\ell \;\;
\sum_{p=0, \pm 1} C^{JM_J}_{\ell m_z 1 p} \ket{\ell m_z} \otimes \ket{1 p},
\ee
where here we have used
$p =0, \pm 1$ to denote the polarization. The eigenfunction
$\phi^a$ of \eq{Nphi} can then be constructed with the matrix elements,
\be
(\phi^a)_{n_1 n_2} = (\langle n_1 n_2 | \otimes \bra{a}) | J M_J \rangle ,
\ee
where  $\ket{ n_1 n_2 }$ are the basis states of the matrix space and
$\ket{a}$ are the basis  states of the 3-dimensional vector space. As a result,
we obtain the eigenvector $\phi^a$ as given by the vector spherical harmonics
\be
\phi^a =  \sum_{m_z=-\ell}^\ell \;\;
\sum_{p=0, \pm 1} C^{JM_J}_{\ell m_z 1 p} \; e^p_a \; \Yh^\ell_{m_z} :
= (\Yh^{JM_J}_\ell)_a,
\ee
where $e^p_a := \langle a| 1p \rangle$ is the $a$-th component
  of the polarization
  vector $e^p$. 
Note that the eigenvalue \eq{eee} is independent of $M_J$ and so each of the
eigenvalues in \eq{eee} has a degeneracy of $2 J +1$.  In total, we have
$3N^2$ eigenmodes in \eq{eee}. Not all of these modes are admissible
fluctuations, however. 

The perturbation is unstable if the eigenvalue is negative, i.e.
for the modes $\ell =0, J =1$ and $\ell =1,
J=2$.  The mode $\ell =1, J=2$ corresponds to a scaling $\d X^a
\propto X^a$ of the fuzzy sphere. This changes the energy of the fuzzy sphere.
The mode $\ell =0,
J =1$  corresponds to a shift of an overall $U(1)$ factor, which
is not allowed since $X^a$ is traceless. Fluctuations
represented by the other modes, energy preserving or not, are
stable. Therefore we find that the single fuzzy sphere solution is
stable against linearized energy preserving perturbations.

Finally, we note that upon quantization, the quadratic fluctuation modes
become oscillators with frequency determined by $\l$ in \eq{eee}. We adopt a
definition of the quantized bosonic Hamiltonian by subtracting a constant
such that the fuzzy sphere has vanishing zero point energy. In other word,
the ground state of the single fuzzy sphere has
a vanishing zero point energy for both
the bosonic as well as the fermionic oscillations.

\section{Quantum black hole as fuzzy sphere with a half-filled Fermi sea}

Consider the fuzzy sphere solution with a half-filled Fermi sea ($n=0$). 
The total energy of the fuzzy sphere system is given by
\be \label{E-half-0}
E=  \frac{\g N M_P}{2}. \quad \mbox{where $\g := a_2 +2a_3  -1$}. 
\ee
To compare with the energy of the Schwarzschild black hole, let us introduce 
dimensional coordinates $Y^a = 2l_P X^a$ for some length scale $l_P$, the
fuzzy sphere solution becomes
\be
   [Y^a, Y^b] = \frac{2 i R}{\sqrt{N^2-1}}\e_{abc} Y^c,
   \quad \sum_a Y^{a2} = R^2 \id,
   \ee
   where $R^2 = (N^2-1) l_P^2$.
This describes a fuzzy sphere of radius $R = N l_P$ in the large $N$ limit.
The energy \eq{E-half-0} can then be put in the form
\be \label{ER}
E = \frac{R}{2G}
\ee
if we identify $M_P$ and $l_P$ with the Newton constant $G$ as
\be \label{Gc}
G = \frac{l_P}{\g M_P}.
\ee
The relation \eq{ER} is precisely the Schwarzschild mass-radius relation if
the fuzzy sphere is identified with the black hole
horizon and the total energy
of the system is identified with the mass $M$ of the black hole 
\be \label {EM}
E =M.
\ee
We note that the relation \eq{EM} holds
if we assume the principle of equivalence of internal
(non-gravitational) energy
and gravitational mass.
In the next section, we will
consider multiple fuzzy spheres configuration in our model and show that the
interaction energy depends of the product $G M_1 M_2$ precisely as in 
Newton's gravity
if the gravitational mass given by \eq{EM}. 
This gives a simple explanation of the equivalence principle from the quantum
mechanics.

Next let us consider the microstates counting. The level $n$
eigenvalue has a degeneracy of
\be
\Omega_n =   \binom{2N^2}{N^2 +n},
\ee
which corresponds to the ways to fill exactly $N^2+n$ of the oscillator
levels.
For the $n=0$ energy state, we have
\be
\Omega_0 = 2^{2N^2}
\ee
in the leading order of large $N$. 
These microstates of the system
at the energy \eq{ER} give rises to the entropy
$S= \log_2 \Omega_0$: 
\be \label{S-QM}
S = 2N^2 .
\ee
This is precisely the Bekenstein-Hawking entropy
of a Schwarzschild black hole if
\be \label{lP}
l_P = \sqrt{\frac{2G}{\pi}}. 
\ee
As a result of \eq{Gc}, $M_P$ is given by
\be
M_P = \frac{1}{\g} \sqrt{\frac{2}{\pi G}}. 
\ee
It is remarkable that 
our quantum mechanical theory \eq{H} admits solution that matches precisely
the desired properties of a quantum black hole.
We thus propose that a quantum black hole is described in our theory
by a fuzzy sphere geometry with a half-filled fermi sea.
We note that the matching works for any coefficients $a_0, a_2, a_3$
as long as $\g>0$.
This arbitrariness of the action
can be fixed by considering the rotating Kerr black hole
\cite{Chu:2024edh}. 
We note that the state $n=0$ is singled out by the fact that
$\Omega_n$ is maximized at $n=0$. It is amazing that a quantum black hole
in our description is  characterized to be one,
given the mass of the black hole is fixed,
having the maximal allowed amount of microstates.
In the next section, we provide further
justification by showing that the fuzzy spheres interact with each other
with a $G M_1 M_2$ dependence that is characteristic  of
Newton gravity.

Finally let us comment on
the Bekenstein entropy bound \cite{Bekenstein:1980jp}
which set the maximal  amount of entropy
that can be contained within a region of space with radius $R$ and energy $E$:
\be
S \leq 2\pi RE.
\ee
Although counter examples of this bound is known, it has been proven
in quantum field theory \cite{Casini:2008cr}.  Presumably it also holds in a
consistent theory of quantum gravity. Let us check it against our theory.
Indeed the bound is saturated for the $n=0$ state of the fuzzy
sphere. For positive $n$, the inequality is satisfied since $S(n) <
S(0)=2\pi R E_0 <2 \pi R E_n$.  However, it is easy to see that the
bound is violated for negative $n$. Physically, it means the fuzzy
spheres in $n=0$ state (black hole) and the
positive $n$ more energetic  state
(presumably represent stars and compact objects) are allowed in the
theory, while those states with negative $n$ should be excluded. It is
possible that these states becomes unstable in the higher order
perturbation theory. It would be interesting to understand this
better.

\section{Multiple black holes}

The reproduction of the static properties of a quantum black hole  supports
the conjecture that  the fuzzy sphere solution does describe
a quantum black hole and that the theory \eq{H}
is a creditable proposal. Below we
perform a preliminary analysis on the interaction of
fuzzy spheres and show that  some characteristic features of Newton gravity
can be reproduced.

Let us consider block diagonal configuration of the following form
\be\label{block-X}
X^a =
\left(
\begin{array}{c|c}
X_1^a & 0 \\
\hline
0 & X^a_2 
\end{array}
\right),
\ee
where $X_\ri^a$ is of dimensions $N_\ri \times N_\ri$, $\ri =1,2$ and $N=N_1+N_2$.
We consider large $N_\ri$
such that the ratios $\a_\ri:= N_\ri/N$ are fixed. The matrices $X^a$ are
traceless, but  $X^a_\ri$ are generally not. 
Let us introduce the coordinates
\be
x^a_\ri := \frac{1}{N_\ri} \tr X_\ri^a,
\ee
which satisfies
\be \label{com}
N_1 x_1^a + N_2 x_2^a =0,
\ee
due to the tracelessness of $X^a$.
We note that since the mass of the isolated black hole is  given by
\be
M_\ri = \frac{\g N_\ri M_P}{2},
\ee
the relation \eq{com} allows $x^a_\ri$ to be interpreted as
the location of the black
holes with respect to the center of mass of the system.
As a result, $X^a_\ri$ can be decomposed into it's
trace and the traceless part $X_\ri^{0a}$ as
\be
X^a_\ri = x_\ri^a + X_\ri^{0 a}
\ee
and
the Hamiltonian of the theory is given by
\be \label{H-block}
H = H_1 + H_2 + H_{12},
\ee
plus kinetic terms for $X^{0a}_\ri$ and $x_\ri^a$.
Here
\be
H_\ri := H_\ri^0 +   H_\ri', \quad \ri =1,2
\ee
where the term
\be \label{HBF}
H_\ri^0 := \a_\ri^2 
 \tr_\ri \left[
-  \left(\frac{M_P}{N_\ri^2} [X_\ri^{0a},X_\ri^{0b}]^2
+ \frac{4M_P}{N_\ri^2} X_\ri^{0a2}\right)
+ \frac{a_2 M_P}{N_\ri^2} \mathop{:}  \psi^\dag \s^a X_\ri^{0a}  \psi\mathop{:} 
\right] +a_3 r_{X_\ri} M_P
\ee
is obtained from the non-abelian part $X_\ri^{0a}$, while the fermionic term
\be
H'_\ri  = \frac{a_2 M_P}{N^2} 
\sum_{m,n =1}^{N_\ri}
\mathop{:} \psi^\dag_{mn} \s^a x_\ri^a  \psi_{mn}\mathop{:} 
\ee
and the bosonic term
\be
H_{12} := -\frac{4 M_P}{N^2}(N_1 x_1^2+ N_2 x_2^2)
\ee
are obtained from the abelian $U(1)$ part $x^a_\ri$  of the configuration.
In the above,
the trace $\tr_\ri$ is taken over the $N_\ri \times N_\ri$ subspace.
Finally 
the terms $H_\ri', H_{12}$ represent
interaction between the two blocks and depend on their separation distance
(in dimensionless matrix units)
\be \label{x12}
\D x := \sqrt{(x_1^a-x_2^a)^2}.
\ee
This comes because \eq{com} implies that
$x_1^a = \a_2 (x_1^a-x_2^a)$ etc. As a result,
the fermionic interaction term $H'_\ri$ is proportional to $\D x$,
and the bosonic interaction term is  proportional to $\D x^2$
\be
H_{12} = - \frac{4M_P}{N} \a_1 \a_2 \D x^2.
\ee

In general,  solutions to the equation of motion of \eq{H-block}
are not static.
Nevertheless, one can consider static configuration and use it to derive
the static potential between the blocks. More general velocity dependent
forces  can also be derived by considering time dependent configuration.
Therefore let us consider the equation of motion in the static limit.
Note that $H^0_\ri$  is identical to the Hamiltonian for a single
black hole except for a replacement of $M_P \to \a_\ri^2 M_P$. 
As a result, we obtain the same equation of motion and  $X^{0a}_\ri$ is solved by
fuzzy sphere with radius
\be
R_\ri = N_\ri l_P.
\ee
The Hamiltonian $H_\ri^0$ admits the eigenstates
\be \label{es-H}
\left|\Psi^{p_1 \cdots p_r q_1 \cdots q_s}_{k_1 \cdots k_r q_1 \cdots q_s}\right\rangle
:=\xi^{p_1\dag}_{k_1}\cdots \xi^{p_r\dag}_{k_r} \chi^{q_1\dag}_{l_1} \cdots
\chi^{q_s\dag}_{l_s} \ket{0},
\ee
where $\xi^{p}_k, \chi^{p}_l$ are defined as in \eq{xi-chi}
using the
eigendecomposition \eq{K-eigen} for $\s^a X_\ri^{0a}$ and the indices
$m,n$ range over the block of $X_\ri$.
Consider the set of states
where half of the oscillators
over the fuzzy sphere are excited (i.e. $r+s = N_\ri^2$)
the fuzzy sphere has the energy
\be \label{H-unperturbed}
H^0_\ri = \frac{M_P N_\ri}{2} (a_2+ 2 \a_\ri a_3  -\a_\ri^2)
\ee
The coarse graining of this ensemble of microstates
gives the Bekenstein-Hawking entropy $S_\ri = 2N_\ri^2 =A_\ri/4G$.

Next, let us consider the interaction term $H^a_\ri$.
For small $\D x /N$, we can
consider $H^a_\ri$ to be a perturbation and
use the perturbation theory
for degeneracy eigenstates to determine its correction to the energy.
This requires the
knowledge  of the matrix elements
$\langle i | H'_\ri | j \rangle $ and
it's eigenvalues $\lambda_p$. Here
$\ket{i}, i = 1, \cdots, \Omega_0:= 2^{2N_\ri^2}$ are the unperturbed
states \eq{es-H}
with $r+s =N_\ri^2$. This is however not only impossible to do as
it involves too large a number of states, Physically, since
the set of microstates is not observed, it is more meaningful to consider the
average corrections over the set of microstates:
\be
\la H'_\ri \ra
:= \sum_{\ket{i} \in\mathbb{V}_0} p_i
\bra{i} H'_\ri \ket{i}, 
\ee
where $p_i$ is the probability of occurrence of the state $\ket{i}$.
For an isolated system,  we can take the probability to
be equal for each of the state and hence $p_i = 1/\Omega_0$.
This corresponds to a microcanonical ensemble. 
We show in the appendix that this ensemble average is given by
\be
\la H'_\ri \ra = \frac{a_2 M_P}{N^2}\tr( \s^a X_\ri^{0a} \s^b x_\ri^b) =0.
\ee 
As a result, the total energy of the two fuzzy spheres system is given by
\be \label{H-smallx}
E= \mbox{const.}
-\frac{4 a_2 M_P}{N}\a_1 \a_2 \D x^2.
\ee
where the constant term is independent of $\D x^2$.
The second term is a correction to the first term in the leading ordering
of small  $\D x^2/N^2$.
In
general, one should include the higher order perturbation
as well as the interaction
term from the off diagonal blocks of \eq{block-X}
and quantum loop effects \cite{Becker:1997wh}. By summing over all these
contributions, one can expect the final result for the energy of
the two fuzzy spheres system to take the form
\be
E = \mbox{const.} - \g N M_P \a_1\a_2 f(\frac{\D x}{N}),
\ee
where $f$ is some function of $\D x/N$. 
It takes the form $f(x) = 1+ \frac{4a_2}{\g}x^2 + O(x^3) $
for small $x$ in our tree level computation.
Defining the coordinate distance between the two
black holes as $r= \D x \, l_P$,
and define the interaction energy as $V(r):= E(r) - E(r=\infty)$. 
We have
$V(r) = -\frac{G M_1 M_2}{r} g(r)$
where $M_\ri = \g N_\ri M_P/2$ is the mass of the
black hole in isolation, and $g (r) := \frac{4r}{R} f(\frac{r}{R})$.
Note that the factor $G M_1 M_2$ appears exactly
the way one expects for Newton gravity.
In order to reproduce Newton gravity
\be V(r) = - \frac{G M_1 M_2}{r}
\ee
in the large distance limit
$r/R = \D x /N \gg 1$, the function $f$ has to be
such that $g = 1 + O(1/r)$.
In order to check whether Newton gravity arises in the large distance limit, 
it is important to devise method to reliably compute the potential $V(r)$
between the fuzzy spheres.
We note that in our model a classical force arises between the fuzzy
sphere due to the mass term.  It is also possible to consider a
variant of our model by considering a Chern-Simons term instead of a
mass term, see. e.g. \cite{Iso:2001mg,Valtancoli:2002rx,
  Azuma:2004ie}
for reduced matrix model with a Chern-Simons term.
The model also
admits fuzzy spheres as solution but in this case
there is no classical force between them and the Newton gravity would
have to emerge from the loop, similar to the situation in the
BFSS matrix model \cite{
  Taylor:2001vb,Becker:1997wh}.
We leave this important issue for  further analysis.
More generally it is
important to understand whether and how tensorial Einstein gravity
may emerge in the low energy approximation in our model.

\section{Discussion}

In this paper we have proposed a
model of quantum space and gravity
as a large $N$ quantum mechanics of non-abelian bosonic and fermionic
coordinates.  The quantum mechanics has static solution whose bosonic
part is given by a fuzzy sphere. Over the fuzzy sphere geometry, the
fermionic part of the theory is given by a collection of fermionic
oscillators all with the same frequency.  The energy state where
exactly half of the Fermi sea is filled contains the maximal amount of
degeneracy. This half-filled fuzzy sphere observes the
mass-radius relation of a Schwarzschild black hole if the fuzzy sphere
radius is identified with the horizon size. Moreover, the coarse
graining of the set of quantum states in the Fermi sea gives precisely
the Bekenstein-Hawking entropy. As a result, we propose that a quantum
black hole is described by a half-filled fuzzy sphere in our model.

We have also considered interaction between these fuzzy spheres by
including the quantum mechanical perturbation arisen from a separation of
them in the matrix space. We find that the interaction energy between
the fuzzy spheres has a dependence on the masses and Newton constant
exactly as in the Newton gravity.  We conjecture that when the full large
$N$ resummation of all the higher order
quantum effects including
the off-diagonal block terms is performed, the exact result will
capture the correct gravitational interaction between the static
quantum black holes, with the Newton gravity reproduced in the large
distance limit.
The obtained modified gravity may be relevant for the dark matter
and dark energy problem.

In this paper we have considered static solutions.  It would be
interesting to consider non-static solution in order to capture the
relativistic effects.  Our proposed quantum mechanics is for quantum
space. Rotational black hole should be included in our
model. It is interesting to understand whether and how matter would arise
in our quantum mechanics. In particular, how gauge interaction can be
incorporated and how a charged back hole can be obtained.
It is also interesting to consider solution with nontrivial fermionic
background 
so as to arrive at a flat solution.

We note that the change of coordinates $x^\mu \to x'{}^\mu (x)$
in general relativity is replaced in our description by
the unitary transformation
\be
X^i \to U X^i U^{-1}.
\ee
It is important to understand how the diffeomorphism symmetry of
general relativity emerges from the quantum mechanics in some
classical limit, and how
Einstein gravity
emerges and get modified in our theory.
Einstein has believed that geodesic equation should not be taken as an
independent assumption but derived from the field equations for empty
space.  This program can be implemented using our quantum mechanics.
For example, one may consider
the motion of a light probe block in the presence of a heavy one
($N_1 \ll N_2$). In some limit, we may ignore the back reaction of the
light block on the heavy one and take the heavy block as a
background. It is possible that one may encode the
effect of the heavy block on the probe in terms of some  effective
metric and  effective energy-momentum tensor,
and the motion of the probe in the background metric
in terms of some effective geodesic motion.

In the general analysis of AMPS \cite{Almheiri:2012rt}  concerning firewalls
and complementarity, it was concluded that the following three
statements cannot all be true: (i) Hawking radiation is in a pure
state, (ii) the information carried by the radiation is emitted from
the region near the horizon, with low energy effective field theory
valid beyond some microscopic distance from the horizon, and (iii) the
infalling observer encounters nothing unusual at the horizon.  Our
study of the two fuzzy sphere systems show that nothing singular
occurs as the distance between them decreases to zero. Taking one of
them as a probe, it means no fire wall and no black hole
singularity. On the other hand, since the black hole horizon is
described by a noncommutative fuzzy geometry in our model, the
existence of a low energy effective theory is questionable due to
UV/IR mixing \cite{Minwalla:1999px} (see also  UV/IR mixing in fuzzy sphere
\cite{Chu:2001xi}) and modified causality and
non-locality of noncommutative field theory
\cite{Chu:2005nb}. It would be interesting to study these points further.
In our description, the fuzzy sphere is dynamical in general and it will
be interesting to develop some kind of effective membrane description
for the horizon, and to compare  it with the classical
membrane paradigm \cite{Thorne:1986iy}.

\section*{Acknowledgments}
We thank Andrew Cohen, Pei-Ming Ho,  Hikaru Kawai and Harold Steinacker
for interesting discussion.
The support of this work by NCTS and
the grants 110-2112-M-007-015-MY3 and
113-2112-M-007-039-MY3 of the National 
Science and Technology Council of Taiwan is gratefully acknowledged. 

\appendix

\section{Ensemble average of 1st order perturbation}
In this appendix, we use perturbation theory to compute the average correction
to the energy of the fuzzy spheres system due to the fermionic term of the
Hamiltonian.

In general,
consider adjoint fermions $\psi_{mn}$ of $SU(N)$ and the Hamiltonian
\be
h_0 =  \mathop{:} \psi^\dag \s^a X^{0a} \psi \mathop{:} ,
\ee
where $X^{0a}$ is given by a fuzzy sphere.
In the large $N$ limit, the matrix $\s^a X^{0a}$
admits the eigenvectors $\cU^{p}, \cV^{p}$ that corresponds
to positive and negative eigenvalues $N/2, -N/2$ respectively. One has
the eigen-decomposition
\be \label{eigen-X}
(\s^a X^{0a})_{(m\a) (n\b)} =  \frac{N}{2} \sum_{p=1}^{N}
\left(\cU^{p}_{m\a} \cU^{p\dag}_{n\b} - \cV^{p}_{m\a} \cV^{ p \dag}_{n\b}
\right).
\ee
The eigenvectors   $\cU^p, \cV^p$ satisfy the orthonormality condition
  \be
\cU^{p\dag}_{m\a} \cU^q_{m\a} = \cV^{p\dag}_{m\a} \cV^q_{m\a} = \d^{pq}, \qquad 
\cU^{p\dag}_{m\a} \cV^q_{m\a} =0
\ee
and the completeness relation
\be
\cU^{p}_{m\a} \cU^{p\dag}_{n\b} +\cV^{p}_{m\a} \cV^{p\dag}_{n\b} =
\d_{mn} \d_{\a\b}.
\ee
Introducing the fermionic oscillators 
  \be \label{xi-chi-a}
  \xi_k^{p} := \cU^{p \dag}_{n\b} \psi_{n k \b}, \quad
   \chi_k^{p\dag} := \cV^{p \dag}_{n\b} \psi_{n k\b},
   \ee
   where the normal ordering is defined with respect to, then
\be
h_0 = \frac{N}{2}
\sum_{p,k=1}^N 
(\xi_k^{p\dag} \xi_k^{p}+ \chi_k^{p\dag} \chi_k^{p}) .
\ee
As a result, $h_0$ has the eigenstates
\be \label{es-n}
\left|\Psi^{p_1 \cdots p_r q_1 \cdots q_s}_{k_1 \cdots k_r l_1 \cdots l_s}\right\rangle
:=\xi^{p_1\dag}_{k_1}\cdots \xi^{p_r\dag}_{k_r} \chi^{q_1\dag}_{l_1} \cdots
\chi^{q_s\dag}_{l_s} \ket{0},
\ee
where $\ket{0}$ is the Fock vacuum and the eigenvalues
\be
h_0 = n+N^2,
\ee
where $n:= r+s -N^2 = -N^2, \cdots, 0, \cdots N^2$ is the energy level
within the Fermi sea.
The energy level $n$ has a degeneracy
of $\Omega_n = \binom{2 N^2}{n+ N^2}$. Denoting this set of states by
\be
\mathbb{V}_n:= \{\ket{i}, i= 1, \cdots, \Omega_n\},
\ee
where $\ket{i}$
are those states \eq{es-n} with $r+s =n + N^2$. 
In the main body of the text,
we
have identified the set $\Omega_0$ of states with the microstates of the
Schwarzschild black hole. 
Now let us consider the Hamiltonian $h= h_0 + h'$ with an operator $h'$  of
the form
\be
h' :=  
\sum_{m,n =1}^{N} \mathop{:} \psi^\dag_{mn} \tilde{h}'\psi_{mn}\mathop{:} .
\ee
 We are interested in the change of energy of the fuzzy spheres system.

For generality, we will treat $h'$ as a perturbation. Obviously, each of the
 microstates
 $\ket{i} \in \mathbb{V}_0$ will receive a different correction.
 Suppose the individual microstate is not observed,  e.g. as in the case
 of black hole, then
rather than trying to determine the splitting  for each individual states,
it is more meaningful to compute some kind of
average correction over all
microstates of the level $n=0$. As the system is in isolation, we have a
microcanonical ensemble where probability of occurance is equally likely for
all the microstates, i.e. $p_i = 1/\Omega_0$. As a result, what is observed
physically is the ensemble average 
\be
\la h' \ra
:= \frac{1}{\Omega_0} \sum_{\ket{i} \in\mathbb{V}_0}
\bra{i} h' \ket{i}.
\ee
In the main text of the paper, we have claimed that $\la h' \ra =0$ for
the displaced fuzzy spheres system. To see this, let us use
\be
\psi_{mn\a} =\cU^p_{m\a} \xi^p_n + \cV^p_{m\a} \chi^{p\dag}_n
\ee
to rewrite $h'$ as 
\be
h' =  P^{pq} \xi^{p \dag}_k \xi^q_k - Q^{pq} \chi^{p\dag}_k \chi^{q}_k +
T^{pq}\xi^{p \dag}_k \chi^{q \dag}_k + {\rm h.c.}
\ee
where the matrices $P^{pq}$ etc are given by
\bea \label{PQ}
& P^{pq} := \cU^{p\dag}_{m\a} \cU^q_{m\b} \tilde{h}'_{\a\b}, \quad
Q^{pq} := \cV^{p\dag}_{m\a} \cV^q_{m\b} \tilde{h}'_{\a\b}, \quad
& T^{pq} := \cU^{p\dag}_{m\a} \cV^q_{m\b}\tilde{h}'_{\a\b}.
\eea
It is easy to see that the $T$ terms do not contribute to the matrix element.
Also, we have the matrix elements
\be
\bra{i} \xi^{p \dag}_k \xi^q_k \ket{i} =  n^\xi_i \d^{pq},\quad
\bra{i} \chi^{p \dag}_k \chi^q_k \ket{i} = n^\chi_i \d^{pq},
\ee
where
\be
n^\xi_i =  N \binom{N^2-1}{r}\binom{N^2}{s}, \quad
 n^\chi_i =N \binom{N^2-1}{s}\binom{N^2}{r}
 \ee
 for a state $\ket{i}$ of the form \eq{es-n}.
Therefore
\be \label{eval-PQ}
\bra{i} h' \ket{i} = n^\xi_i  \tr P - n^\chi_i \tr Q.
\ee
Using $\sum_r \binom{N^2-1}{r}\binom{N^2}{N^2-r} = \Omega_0/2$, the sum over
all states of form \eq{es-n} with a fixed level $n = 0$
is given by 
\be \label{hPQ}
\sum_{\ket{i} \in\mathbb{V}_0} \bra{i} h' \ket{i}= \frac{N}{2}\Omega_0
\tr (P -  Q).
\ee
Using the definitions \eq{PQ} and \eq{eigen-X}, we obtain
\be \label{h'}
\la h' \ra = \tr(K \tilde{h}'),
\ee
where $K = \s^a X_0^a$ is the kernel for the unperturbed Hamiltonian $h_0$.
We emphasis that the result \eq{h'} holds true in  general
and is independent of the form of $\tilde{h}'$. 
In this paper, we consider a perturbation with the kernel
\be \label{h'-cm}
\tilde{h}' = \s^a x^a
\ee
due to a separation of the fuzzy spheres. As a result, $\la h' \ra =0$
 since $\tr(X^{0a}) =0$. We note that this result is also expected since 
 the perturbation \eq{h'-cm}
 takes the form of a spin in an external magnetic field
 $\vec{x}$. As  the fuzzy sphere on which the microstates are defined 
 is isotropic, the magnetic energy averaged
 over the set of microstates
 is zero.

\bibliographystyle{utphys}
\bibliography{references}

\end{document}